\documentclass{svjour3}                     
\smartqed  

\usepackage{graphicx}
\usepackage{amsfonts}

\journalname{Few-Body Syst.}

\begin{document}

\title{Universal prohibited zones in the coordinate space of few-body systems}

\author{
 C. G. Bao$^*$ \thanks{* Email: stsbcg@mail.sysu.edu.cn}
}

\institute{State Key Laboratory of Optoelectronic Materials and
Technologies, School of Physics and Engineering, Sun Yat-Sen
University, Guangzhou, 510275, P.R. China}

\date{Received: date / Accepted: date}

\maketitle

\begin{abstract}
In some special zones of the high-dimensional coordinate space of few-body
systems with identical particles, the operation of an element (or a product
of elements) of the symmetry groups of the Hamiltonian on a quantum state
might be equivalent to the operation of another element. Making use of the
matrix representations of the groups, the equivalence leads to a set of
homogeneous linear equations imposing on the wave functions. When the matrix
of these equations is non-degenerate, the wave functions will appear as
nodal surfaces in these zones. In this case, these zones are prohibited. In
this paper, tightly bound 4-boson systems with three types of interaction
have been studied analytically and numerically. The existence of the
universal prohibited zones has been revealed, and their decisive effect on
the structures of the eigenstates is demonstrated.
\end{abstract}


\section{Introduction}

Since various systems may be governed by the same fundamental
law, universality exists in nature. The early indications of
universality in quantum mechanic three-body systems was
discovered before 1970.\cite{r_X1,r_X2,r_X3} The first strong
evidence was given by Efimov in 1970 on three weakly bound
identical bosons.\cite{r_EV1,r_EV2} \ He found that, when the
scattering length is sufficiently long, a sequence of loosely
bound states, the "Efimov states", will appear. The properties
of these states are governed by a universal law, not depend on
the dynamic details of the 3-boson systems. In this paper,
another kind of universality that exists in tightly bound
quantum mechanic few-body systems is revealed and confirmed
numerically.

The Hamiltonian of identical particles is invariant under the
operations of a set of symmetry groups $G_{\alpha }$, $G_{\beta
}$, $\cdots$ (including the permutation
group).\cite{r_RDE,r_RG,r_LEM,r_EJP} Consequently, the
eigenstates $\{\Psi _{i}(X)\}$ are classified according to the
representations of these groups, where $X$ denotes a set of
coordinates and $i$ is a serial number. Let $g_{\alpha }$ be an
element of $G_{\alpha }$, $g_{\beta }$ be that of $G_{\beta }$,
and $\Xi $ denotes a special zone in the high-dimensional
coordinate space. When $X\in \Xi $, the effects of $g_{\alpha}$
and $g_{\beta }$ might be equivalent so that $g_{\alpha }\Psi
_{i}(X)=g_{\beta }\Psi _{i}(X)$. For an example, when $\Xi $ is
the zone of the squares (SQ), $g_{\alpha }$ is a rotation about
the normal of the SQ by $2\pi /4$, and $g_{\beta }$ is a cyclic
permutation of particles, then $g_{\alpha }$ and $g_{\beta }$
are equivalent in $\Xi $. Making use of the representations of
groups, the equivalence leads to a set of homogeneous linear
equations
\begin{eqnarray}
 \sum_{i'}
 [ D_{i'i}^{\alpha }(g_{\alpha })
  -D_{i'i}^{\beta }(g_{\beta }) ]
 \Psi _{i'}(X)
  =  0, \ \ \ (X\in \Xi ),
 \label{e01_sum}
\end{eqnarray}
where $D_{i'i}^{\alpha }(g_{\alpha })$ are the matrix elements of
the representation. When the matrix of this set of equations is
non-degenerate, the set $\Psi _{i'}(X)$ must be zero in $\Xi $. In
this case, $\Xi $ becomes a prohibited zone (PZ) and the wave
function appears as an inherent nodal surface (INS).\cite{r_BCG1} \
Eq.~(\ref{e01_sum}) demonstrates that, for each pair of equivalent
operations, a constraint will be imposed on the eigenstates. Since
the matrixes of representations are absolutely irrelevant to
dynamics, the constraint is universal\ disregarding the kind of
systems (nuclear, atomic, or molecular) and the details of dynamic
parameters. It implies that the states of different systems but
belonging to the same set of representations will have exactly the
same PZ, and their wave functions will have exactly the same INS. On
the other hand, some zones are important to binding (say, for 4-body
systems, the zones of equilateral tetrahedron (ET) and SQ). Whether
these zones are prohibited is crucial to the binding energy and the
geometric character of a state. Furthermore, the number and the
locations of the nodal surfaces in a wave function in general
determine the strength and mode of oscillation. In particular, the
existence of the INS implies an inherent mode. Thus, the eigenstates
would be seriously affected by the universal symmetry constraint.

The decisive effect of the symmetry constraint on the triply and
quadruple excited intrashell states of atoms has been revealed
previously.\cite{r_BCG2,r_BCG3,r_BCG4,r_BCG5,r_BCG6} Accordingly,
these states can be naturally classified according to their inherent
nodal structures.\cite{r_MT,r_PMD} For 4-boson systems, a number of
predictions on the structures and internal modes of oscillation have
been made previously.\cite{r_BCG7,r_BCG8} However, these predictions
have not yet been confirmed numerically. The present paper
generalizes the work of [18,19] in the following aspects: (i)
Instead of free 4-boson systems, trapped 4-boson systems are
considered. Thereby a number of tightly bound states can be obtained
which are necessary for a systematic analysis.\ \ (ii) In addition
to theoretical analysis, numerical calculations have been performed
so that the effect of symmetry constraint can be appraised
quantitatively. (iii) Three types of interactions have been adopted.
The aim is to demonstrate the similarity among different kinds of
systems.

In the next section, the symmetry constraints imposing on
4-boson systems are studied theoretically. We have chosen
appropriate sets of coordinates so that Eq.~(\ref{e01_sum})
appears in very simple forms and the analysis becomes
transparent. Then, an isotropic trap together with three types
of interaction are introduced, and numerical calculations are
performed to diagonalize the Hamiltonian. Under the trap the
total orbital angular momentum $L$, its $Z$-component $M$, and
the parity $\Pi $ are good quantum numbers. Accordingly, an
eigenstate with the c.m. motion removed can be denoted as $\Psi
_{LM,i}^{\Pi }$, where $i$\ denotes the $i$-th state of a
$L^{\Pi }$-series. Mostly the $i=1$ states (the lowest one) are
concerned. Therefore, the label $i$\ is dropped hereafter
(except specified). After obtaining the eigenenergies and the
eigenstates, a number of quantities (the root-mean-square
radius, the one-body densities for the particle distribution,
and the shape-densities) are further calculated. Thereby,
inherent physics can be extracted, and a clear comparison among
different kinds of system can be made. The emphasis is placed
to demonstrate the universality of the PZ and the similarity
among different systems. A short discussion on 4-fermion
systems is also given at the end.

\section{Universal prohibited zones in the coordinate space}

The equilateral tetrahedron (ET) and the square (SQ) are the two
most important geometries. We shall study the symmetry constraint
taking place at the ET, SQ, and their neighborhoods. The zone
associated with the extension-contraction of an ET along one of its
two-fold (three-fold) axis is defined as $Z_{2-2}$ ($Z_{1-3}$).
$Z_{2-2}$ and $Z_{1-3}$ are related to the H-type and K-type of
oscillations, respectively, that exist in tetramers.\cite{r_X4} \
When the tetrahedrons of $Z_{2-2}$ are further twisted about the
two-fold axis, the extended zone is defined as $Z_{\mathrm{twi}}$.

\begin{figure}[tbp]
 \centering \resizebox{0.9\columnwidth}{!}{
 \includegraphics{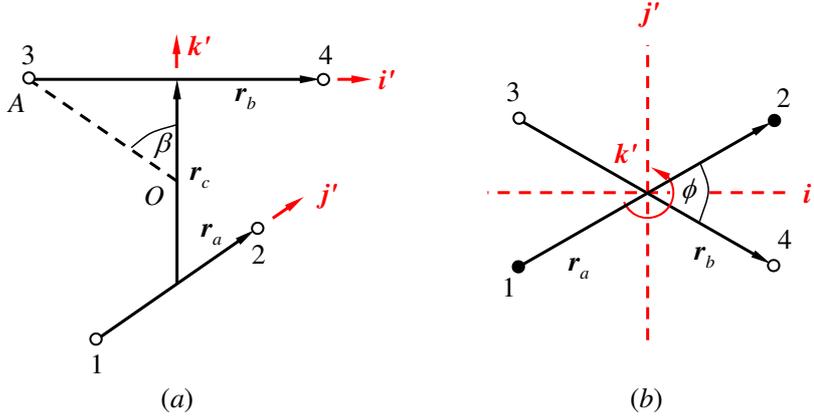} }
 \caption{(a) The zone $Z_{2-2}$ defined in a body-frame with
$\mathbf{r}_c$ lying along the third body-axis $\mathbf{k'}$,
$\mathbf{r}_a$ lying along $\mathbf{j'}$, and $\mathbf{r}_b$
lying along $\mathbf{i'}$. $r_a=r_b$ is assumed. In this zone
the magnitudes $r_a$ and $r_c$ are considered as variable. (b)
The zone $Z_{\mathrm{twi}}$ defined in a body frame with $\mathbf{r}_c $
also lying along $\mathbf{k'}$, $\mathbf{r}_a\perp\mathbf{r}_c$
lying under the $i'$-$j'$ plane, and
$\mathbf{r}_b\perp\mathbf{r}_c$ lying above the $i'$-$j'$
plane. $r_a=r_b$ and $\phi_a=-\phi_b $ are assumed. In this
zone $r_c, r_a=r_b$, and $\phi=\phi_a- \phi_b$ are considered
as variable.}
 \label{fig1}
\end{figure}

When appropriate degrees of freedom are chosen, the general
expression Eq.~(\ref{e01_sum}) could have very simple form. For
$Z_{2-2}$, let the Jacobi coordinates be
$\mathbf{r}_{a}=\mathbf{r}_{2}-\mathbf{r}_{1}$,
$\mathbf{r}_{b}=\mathbf{r}_{4}-\mathbf{r}_{3}$,
$\mathbf{r}_{c}=(\mathbf{r}_{4}+\mathbf{r}_{3}-\mathbf{r}_{2}-\mathbf{r}_{1})/2$,
then $\mathbf{r}_{a}\perp \mathbf{r}_{b}\perp \mathbf{r}_{c}$ and
$r_{a}=r_{b}$ are required. $r_{a}$\ and $r_{c}$ are considered as
variable. We introduce a body frame $\Sigma '$\ with its origin at
the c.m., its $\mathbf{k}'$ lying along $\mathbf{r}_{c}$,
$\mathbf{j}'$ lying along $\mathbf{r}_{a}$, and $\mathbf{i}'$ lying
along $\mathbf{r}_{b}$. $Z_{2-2}$ is shown in Fig.~\ref{fig1}a. This
zone is 2-dimensional in $\Sigma '$.

In general, an eigenstate can be expressed by using the arguments of a body
frame as
\begin{eqnarray}
 \Psi _{LM}^{\Pi }(X)
  =  \sum_{Q}
     D_{QM}^{L}(\Omega )
     \Psi_{LQ}^{\Pi }(X'),
 \label{e02_Psi}
\end{eqnarray}
where $X$ denotes the set of coordinates relative to a fixed frame
$\Sigma $, and $X'$ denotes the set relative to $\Sigma'$, $\Omega$
denotes the set of Euler angles responsible for the transformation
from $\Sigma '$ to $\Sigma $.\cite{r_EAR} Let $P$ denotes an
arbitrary particle permutation, $I$ a space inversion with respect
to the c.m., and $R_{\phi }^{\mathbf{v}}$ a rotation about the axis
$\mathbf{v}$ by the angle $\phi $. In $Z_{2-2}$ we have the
equivalence $IP\doteq R_{\pi /2}^{\mathbf{k}'} $. When $IP$ act at
the left side of Eq.~(\ref{e02_Psi}) and $R_{\pi /2}^{\mathbf{k}'}$\
acts at the right side, we obtain the constraint
\begin{eqnarray}
 (-i)^{Q}
 \Psi _{LQ}^{\Pi }(X')
  =  \Pi
     \Psi_{LQ}^{\Pi }(X'),
 \label{e03_Psi}
\end{eqnarray}
which is a simple form of Eq.~(\ref{e01_sum}). Therefore, $\Psi
_{LQ}^{\Pi }(X')$ is nonzero in $Z_{2-2}$ only if $Q=0,\pm
4,\cdots$ (when $\Pi =1$), or $Q=\pm 2,\pm 6,\cdots$ (when $\Pi
=-1$).

We have also the equivalence $IP\doteq R_{\pi }^{\mathbf{j}'}$,
which leads to the second constraint
\begin{eqnarray}
 (-1)^{L+Q}
 \Psi _{L,-Q}^{\Pi }(X')
  =  \Pi
     \Psi _{LQ}^{\Pi}(X').
 \label{e04_Psi}
\end{eqnarray}
Therefore, the $Q=0$ component exists only if $(-1)^{L}=\Pi $.

With this in mind, we know that $Z_{2-2}$ is a PZ for $0^{-}$
and $1^{-}$ states (because they have only $|Q|\leq 1$
components), and for $1^{+}$ and $3^{+}$ states (because they\
do not have $|Q|\geq 4$ components and their $Q=0$\ component
violates the requirement $(-1)^{L}=\Pi $).

When $r_{c}=r_{a}/\sqrt{2}$, the particles form an ET. The zone
of ET is denoted as $Z_{\mathrm{ET}}$\ which is one-dimensional
in $\Sigma '$ and is a subspace of $Z_{2-2}$.\ In
$Z_{\mathrm{ET}}$ OA (refer to Fig.~\ref{fig1}a) becomes a
three-fold axis of an ET. Thus, additionally, we have
\begin{eqnarray}
 R_{2\pi /3}^{\mathbf{OA}}
  =  R_{-\beta }^{\mathbf{j}'}
     R_{2\pi /3}^{\mathbf{k}'}
     R_{\beta }^{\mathbf{j}'}
     \doteq
     P,  \label{e05_R}
\end{eqnarray}
where $\beta =\cos ^{-1}(\sqrt{1/3})$ is the angle between OA
and $\mathbf{k}'$. This leads to a set of homogeneous linear
equations as
\begin{eqnarray}
 \sum_{Q'}
 [ ( \sum_{Q''}
     d_{Q'Q''}^{L}(-\beta )
     e^{-i\frac{2\pi }{3}Q''}
     d_{Q''Q}^{L}(\beta ) )
  -\delta _{Q'Q} ]
 \Psi_{LQ'}(\mathrm{ET})
  =  0,
 \label{e06_sum}
\end{eqnarray}
where $d_{Q'Q''}^{L}$ is the matrix elements of the rotation about
$\mathbf{j}'$.[21] Under the constraints given by
Eqs.~(\ref{e03_Psi}), (\ref{e04_Psi}) and (\ref{e06_sum}), nonzero
solutions exist at the ET only if $L^{\Pi }=0^{+}$, $3^{-}$ and
$4^{+}$ (if $L\leq 4$). They are called ET-accessible
(ET-\textbf{ac}) states.\cite{r_BCG7,r_BCG8}

When $r_{c}\rightarrow 0$, the shapes in $Z_{2-2}$ tends to a SQ
with its plane lying on the $\mathbf{i}'$-$\mathbf{j}'$ plane
(Fig.~\ref{fig1}a). The zone of SQ is denoted as $D_{\mathrm{SQ}}$\
which is also one-dimensional in $\Sigma '$. For $D_{\mathrm{SQ}}$
we have $I\doteq P$. Thus, the SQ can only exist in $\Pi =+1$
states. As a subspace of $Z_{2-2}$, the SQ must be constrained by
Eqs.~(\ref{e03_Psi}) and (\ref{e04_Psi}). Thus the SQ-\textbf{ac}
states are $0^{+}$, $2^{+}$ and $4^{+}$ (if $L\leq 4$). Among them,
$0^{+}$ and $4^{+}$ are both ET- and SQ-\textbf{ac}.

The geometries in $Z_{1-3}$ are the regular trihedral pyramids.
Let $h$\ denotes the height which is lying along the 3-fold
axis of the pyramid, and $s$\ denotes the side length of the
base regular triangle. $h$ and $s$ are variable. Let
$\mathbf{k}'$\ be lying along the 3-fold axis, and
$\mathbf{j}'$\ along a height of the base triangle. Then we
have $R_{2\pi /3}^{\mathbf{k}'}\doteq P$ and $R_{\pi
}^{\mathbf{i}'}\doteq IP$. Accordingly, $Z_{1-3}$ is a PZ\ for
$0^{-}$, $1^{+}$ and $2^{-}$ states.

In $Z_{\mathrm{twi}}$ (refer to Fig.~\ref{fig1}b where
$\mathbf{r}_{c}$ is lying along $\mathbf{k}'$ and can not be
seen) $\mathbf{r}_{a}\perp \mathbf{r}_{c}$,
$\mathbf{r}_{b}\perp \mathbf{r}_{c}$, $r_{a}=r_{b}$ are
required. $r_{a}$, $r_{c}$ and the angle $\phi $\ between
$\mathbf{r}_{a}$ and $\mathbf{r}_{b}$ are considered as
variable. In this zone $\mathbf{k}'$ is a 2-fold axis, and
$\mathbf{i}'$ lying along the bisector of $\phi $ is also a
2-fold axis. We have $R_{\pi }^{\mathbf{k}'}\doteq P$ and
$R_{\pi}^{\mathbf{i}'}\doteq P$. Accordingly,
$Z_{\mathrm{twi}}$ is a PZ\ for $1^{-} $ and $1^{+}$ states.

For coplanar shapes, let $\mathbf{k}'$ be vertical to the plane
of the shapes. Then, $R_{\pi }^{\mathbf{k}'}\doteq I$ must
hold. We considered the following three cases.

(i) The coplanar shape is invariant under inversion (e.g., a
parallelogram) and, accordingly, $\mathbf{k}'$ is a two-fold
axis. The associated zone is denoted as $D_{A}$, and the
equivalence $I\doteq P$ holds.

(ii) The shape contains at least a two-fold axis on the
$\mathbf{i}'$-$\mathbf{j}'$ plane (say, an isosceles
trapezoid). The zone is denoted as $D_{B}$, and the equivalence
$R_{\pi }^{\mathbf{j}'}\doteq P$ holds.

(iii) The intersection of $D_{A}$ and $D_{B}$, namely,
$D_{A\cap B}$. This zone includes the rectangles and diamonds.
Specifically, $D_{\mathrm{SQ}}\in D_{A\cap B}$.

Associated with the above zones, the PZ\ are summarized in
Table~\ref{tab1}. It is emphasized that the existence of the PZ is
universal disregarding the kinds of 4-boson systems. Since the
prohibition is absolutely not violable, the effect of the PZ is
decisive. The features of all the eigenstates can be thereby more or
less predicted. For examples, $0^{+}$ is free from symmetry
constraint. Thus, it can choose the structure most favorable to
binding, and its wave function is expected to surround an ET.
$2^{+}$ is not ET-\textbf{ac} but is SQ-\textbf{ac}, therefore its
wave function is expected to surround a SQ, and so on.

For the eight zones listed in Table~\ref{tab1}, the most important
zones are the $Z_{\mathrm{ET}}$ and $D_{\mathrm{SQ}}$. The
accessibility of these two zones is crucial to the energies of the
eigenstates as shown below. When a zone is accessible to a state but
some subspaces inside the zone are not (say, $Z_{\mathrm{twi}}$ is
accessible to $0^{-}$ but its subspace $Z_{2-2}$ is not), the zone
is not a stable zone because the PZ inside the zone implies the
existence of inherent nodal surfaces, and therefore specific
oscillations are excited (examples are given below).
\begin{table}[htb]
 \caption{The prohibited zones (PZ) emerge in the coordinate
space for the $L^{\Pi }$ states ($L\leq 4$) of 4-boson systems.
The second column gives the dimension of the zones in the body
frame. The appearance of a PZ is marked by an "\checkmark"
(say, $Z_{\mathrm{twi}}$ is a PZ\ for $1^{+}$ and $1^{-}$\
states).}
  \begin{center}
    \label{tab1}
    \begin{tabular}{c|c|*{10}{c}}
      \hline\hline
       & dim & $0^{+}$ & $1^{+}$ & $2^{+}$ & $3^{+}$ & $4^{+}$ & $0^{-}$ & $1^{-}$ & $2^{-}$ & $3^{-}$ & $4^{-}$   \\
      \hline
       $Z_{\mathrm{twi}}$     & 3 &  & \checkmark &            &            &  &            & \checkmark &            &            &            \\
       $Z_{2-2}$              & 2 &  & \checkmark &            & \checkmark &  & \checkmark & \checkmark &            &            &            \\
       $Z_{\mathrm{ET}}$      & 1 &  & \checkmark & \checkmark & \checkmark &  & \checkmark & \checkmark & \checkmark &            & \checkmark \\
       $D_{\mathrm{SQ}}$      & 1 &  & \checkmark &            & \checkmark &  & \checkmark & \checkmark & \checkmark & \checkmark & \checkmark \\
       $Z_{1-3}$              & 2 &  & \checkmark &            &            &  & \checkmark &            & \checkmark &            &            \\
       $D_{A}$                & 3 &  &            &            &            &  & \checkmark & \checkmark & \checkmark & \checkmark & \checkmark \\
       $D_{B}$                & 3 &  & \checkmark &            &            &  & \checkmark &            &            &            &            \\
       $D_{A\cap B}$          & 2 &  & \checkmark &            &            &  & \checkmark & \checkmark & \checkmark & \checkmark & \checkmark \\
      \hline\hline
    \end{tabular}
  \end{center}
\end{table}

\section{Trapped 4-boson systems with three kinds of
interaction}

We are going to present numerical results to evaluate quantitatively
the consequence of the PZ. It is assumed that the four particles are
confined by an isotropic harmonic trap $\frac{1}{2}m\omega
^{2}r_{i}^{2}$. The trap is introduced just for supporting more
bound states for a systematic analysis, and will not at all affect
the appearance of the PZ found above. $\hbar \omega $ and
$\sqrt{\hbar /m\omega }$ are used as units of energy and length.
With these units three types of interactions are assumed:
$V_{\mathrm{A}}(r)=10(2e^{-(r/1.428)^{2}}-e^{-(r/2.105)^{2}})$,
$V_{\mathrm{B}}(r)=1000e^{-3r}-40/r^{6}$ (when $r\geq 1.2$) and
$V_{\mathrm{B}}(r)=V_{\mathrm{B}}(1.2)$ (when $r<1.2$) , and
$V_{\mathrm{C}}(r)=15$ (when $r\leq 1$) and $V_{\mathrm{C}}(r)=0$
(when $r>1$), respectively, where
$r=|\mathbf{r}_{i}-\mathbf{r}_{j}|$. $V_{\mathrm{A}}$ has a
short-ranged character and was previously used in nuclear physics
for the $\alpha $-particles, $V_{\mathrm{B}}$ belongs to the Van der
Waals type for atoms, while $V_{\mathrm{C}}$ is just a repulsive
hard core potential. In fact, the three interactions are chosen
quite arbitrary, just to show the possible similarity among
different systems.

With the interaction and the trap, the Hamiltonian is
\begin{eqnarray}
 H
  =  \sum_{i}
     (-\frac{1}{2}
       \nabla _{i}^{2}
      +\frac{1}{2}r_{i}^{2})+
     \sum_{i<j}
     V_{J}
     ( |\mathbf{r}_{i}
      -\mathbf{r}_{j}| ),
\end{eqnarray}
where $J=A,B$, or $C$. A set of basis functions is introduced to
diagonalize the Hamiltonian to obtain the spectra and the
eigenstates. The details are given in the Appendix A. Note that
$V_{\mathrm{A}}$ and $V_{\mathrm{B}}$ both contain a minimum. Thus
the total interaction energy would be lower if the particle
separations are appropriate. Therefore the particles will pursue a
better geometry. The ET (all six bonds can be optimized) and the SQ
(four bonds can be optimized) with an appropriate size will be the
first and second choices. However, for $V_{\mathrm{C}}$, the minimum
is not contained, therefore the pursuit to a better geometry is less
anxious.

\begin{figure}[tbp]
 \centering \resizebox{0.9\columnwidth}{!}{
 \includegraphics{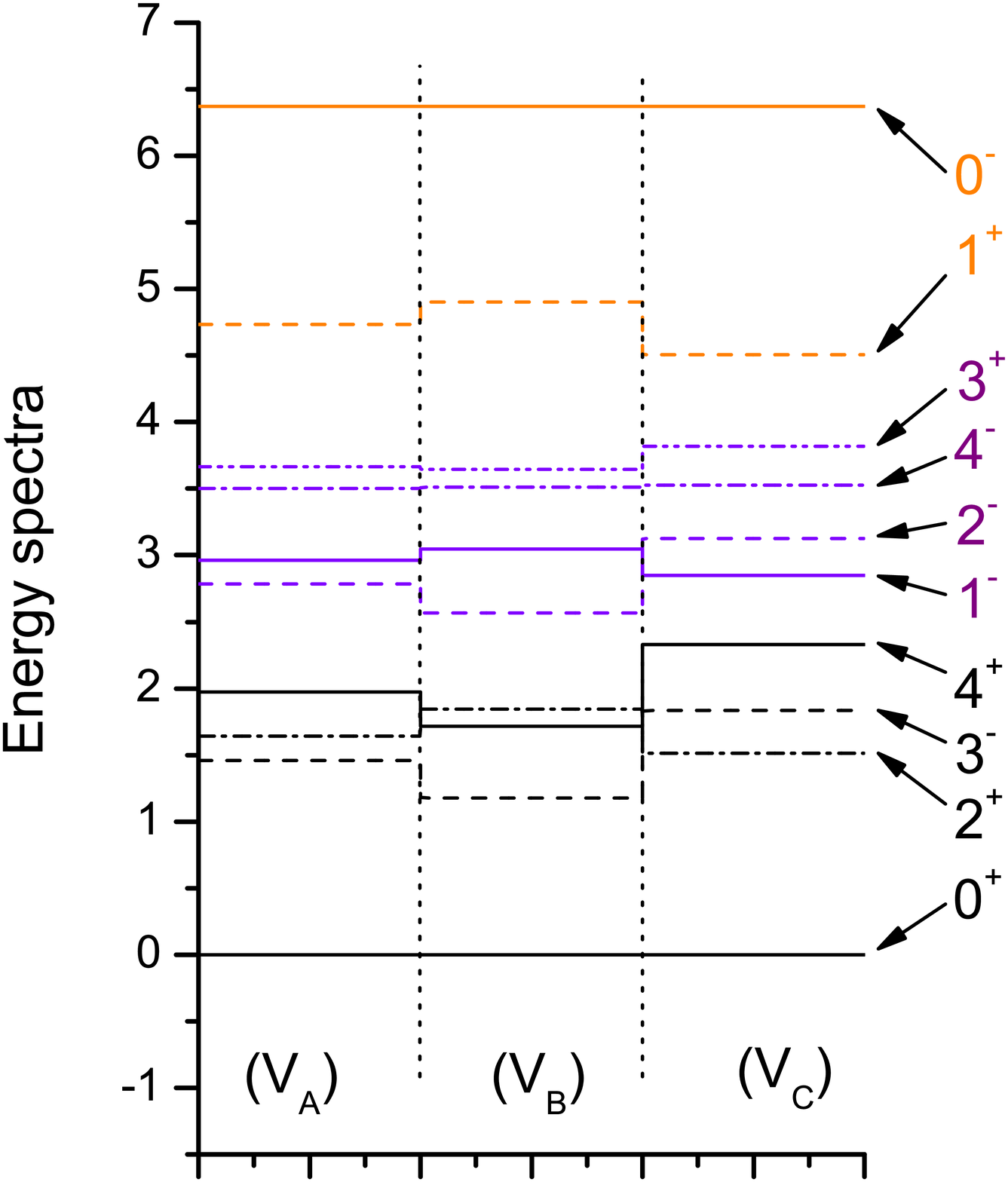} }
 \caption{Spectra of the 4-boson system with $L=0$ to $4$ and
parity $\Pi =\pm 1$. Only the lowest one of a $L^{\Pi }$-series are
given. Three types of interactions $V_{\mathrm{A}}$,
$V_{\mathrm{B}}$, and $V_{\mathrm{C}}$ are considered. The levels
have been shifted so that the three ground states ($0^{+}$) of the
three types have energy zero. The energy unit for $V_{\mathrm{A}}$
is $\hbar \omega $. The energy units for the spectra of
$V_{\mathrm{B}}$ and $V_{\mathrm{C}}$ have been redefined so that
the three $0^{-}$ states have the same value of energy (but in
different units). In all the three spectra, the lowest four belong
to the first group, the highest two belong to the third group. The
four in the middle belong to the second group. The states contained
in the groups are the same for the three types without exception. In
a group the same state is given by the same kind of line for the
three types (say, all the $3^{-}$ states belonging to the first
group are given by the dash-line).}
 \label{fig2}
\end{figure}

\section{Spectra}

The spectra with the three types of interaction are plotted in
Fig.~\ref{fig2}. Although the interactions are very different, the
three spectra have common features. From Table~\ref{tab1} we know
that the $0^{+}$, $2^{+}$, $4^{+}$, and $3^{-} $ are ET-\textbf{ac
}and/or SQ-\textbf{ac}. They form the first group. This group of
states are explicitly lower than the others, and the $0^{+}$ state
is the lowest as expected. Usually, the states with a larger $L$ is
higher due to having a stronger rotation energy. However, for
$V_{\mathrm{A}}$ and $V_{\mathrm{B}}$, $2^{+}$ is higher than
$3^{-}$ because the former is simply SQ-\textbf{ac} while the latter
is ET-\textbf{ac}. Nonetheless, for $V_{C}$, the superiority of ET
over SQ is weaker, thereby the normal sequence in this group is
recovered. For all the three spectra, $0^{-}$ and $1^{+}$ are the
highest because many PZ are contained. These two form the third
group. Although they contain zero or seldom rotation energy, they
are still much higher than the $4^{+}$. This fact demonstrates the
serious effect of the symmetry constraint. The remaining four states
($1^{-}$, $2^{-}$, $4^{-}$, and $3^{+}$) are in the middle, they
form the second group. Totally speaking, the two spectra for
$V_{\mathrm{A}}$ and $V_{\mathrm{B}}$ are very similar (except the
energy scale). The similarity of the spectrum of $V_{C}$\ to the
other two is weaker.

\section{Root mean square radius}

\begin{figure}[tbp]
 \centering \resizebox{0.9\columnwidth}{!}{
 \includegraphics{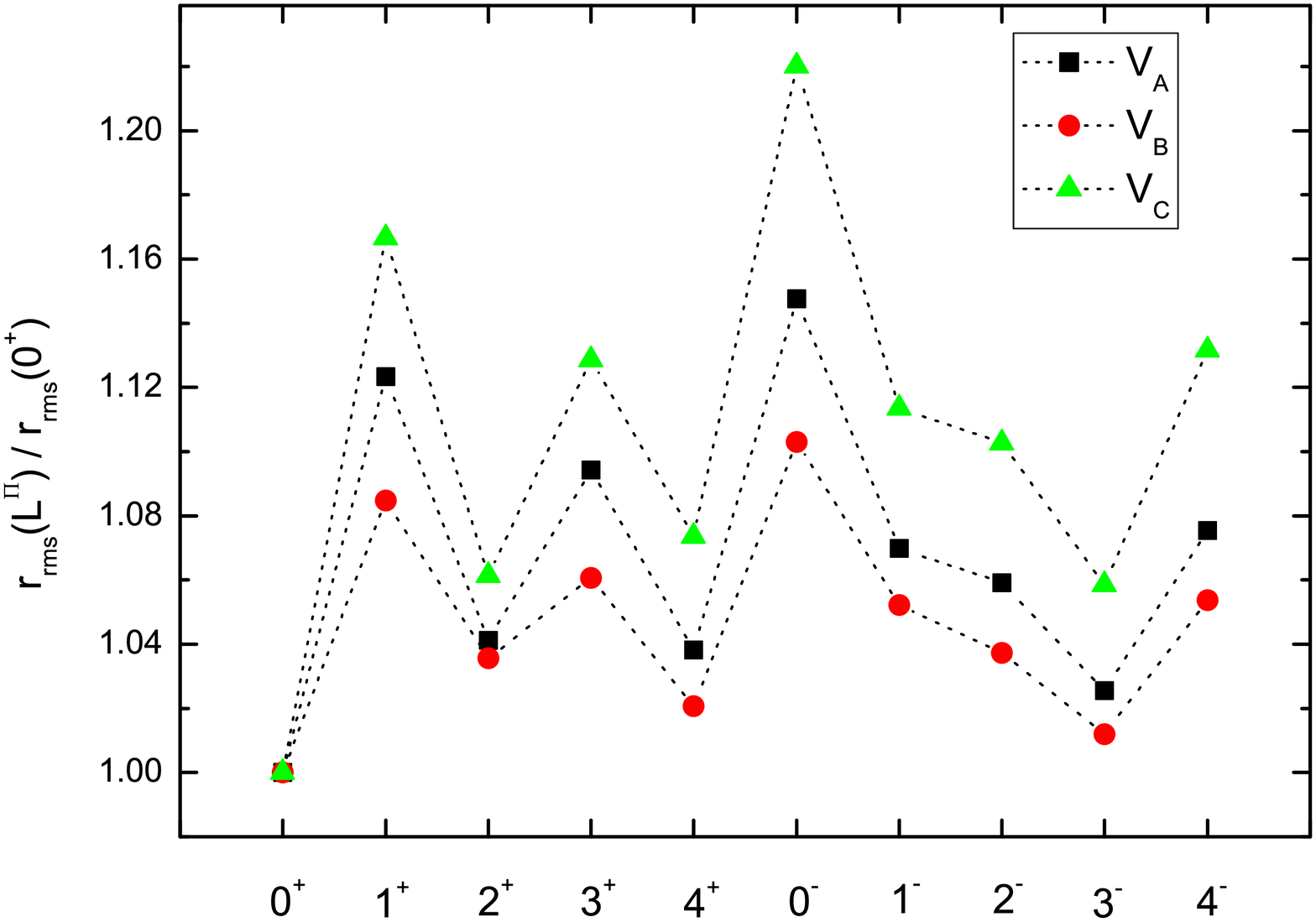} }
 \caption{The ratio
$r_{\mathrm{rms}}(L^{\Pi})/r_{\mathrm{rms}}(0^+)$ for the
$L^{\Pi }$ states of the three types. Square is for the type
$V_{\mathrm{A}}$, circle for $V_{\mathrm{B}}$, and triangle for
$V_{\mathrm{C}}$. }
 \label{fig3}
\end{figure}

We calculate the root mean square radius
$r_{\mathrm{rms}}(L^{\Pi })$ to evaluate the size of each state
(refer to Appendix B). The ratios $r_{\mathrm{rms}}(L^{\Pi
})/r_{\mathrm{rms}}(0^{+})$ are shown in Fig.~\ref{fig3} where
the three dotted lines guiding the eyes go up and down in a
synchronous way. It demonstrates once again the similarity
among different systems. For all the three types of
interactions, the sizes of the ET-\textbf{ac} states $0^{+}$,
$3^{-}$, and $4^{+}$ are relatively smaller, while the sizes of
$0^{-}$ and $1^{+}$ are the largest (because their wave
functions are expelled from many PZ as shown below). If dynamic
effect played an essential role, the largest would have $L=4$
because the particles are pushing out by a stronger centrifugal
force. But in fact not.

\section{Particle distribution}

\begin{figure}[tbp]
 \centering \resizebox{0.9\columnwidth}{!}{
 \includegraphics{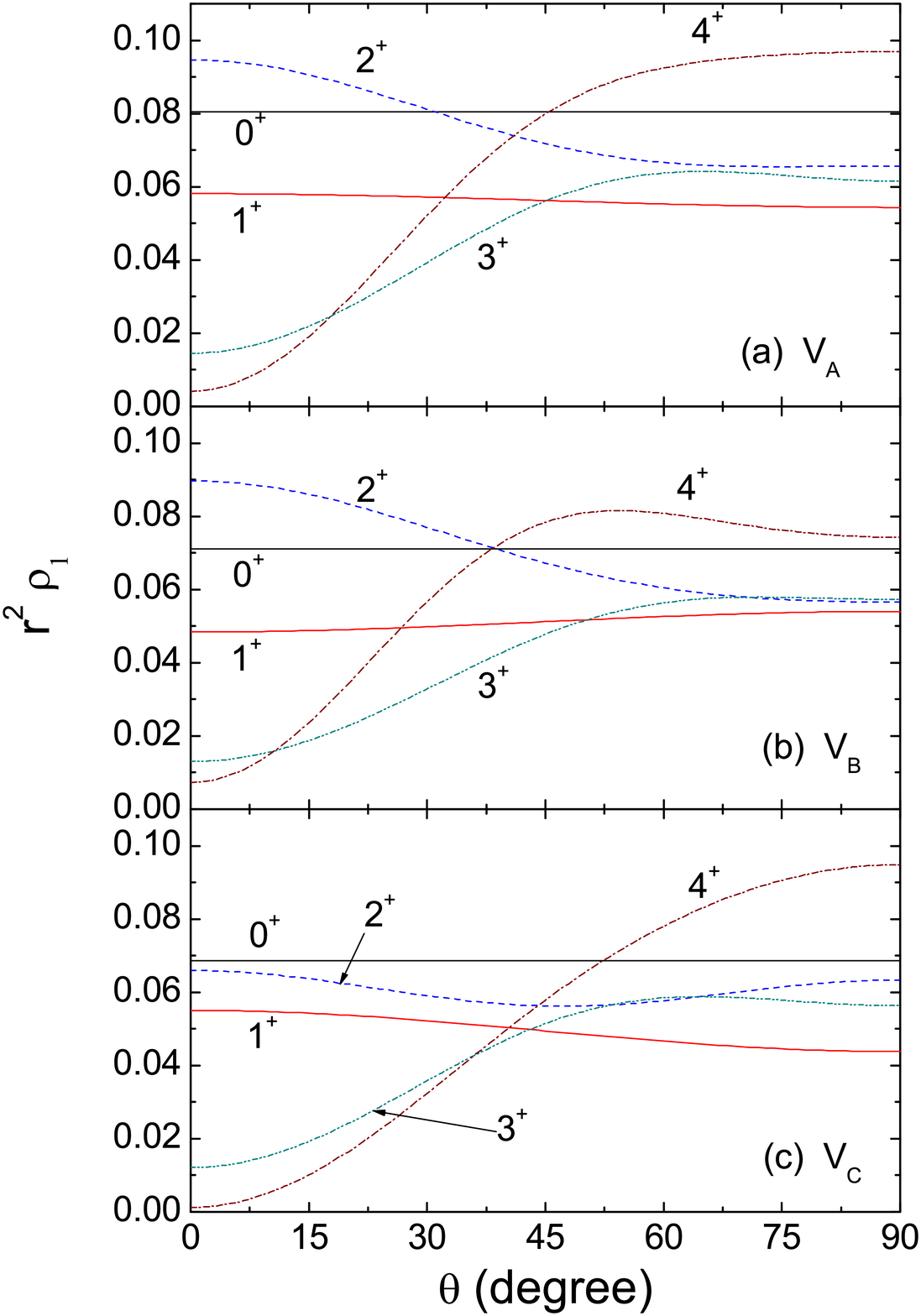} }
 \caption{One-body densities $r^2\rho_1(r,\theta)$ for the even
parity states $\Psi_{LM}^{(+)}$ with $r$ being fixed at
$r_{\mathrm{rms}}(L^{\Pi})$. $M=L$ is chosen. Since
$\rho(r,\theta)=\rho(r,\pi-\theta)$, only $\theta\leq\pi/2$ is
included. }
 \label{fig4}
\end{figure}

\begin{figure}[tbp]
 \centering \resizebox{0.9\columnwidth}{!}{
 \includegraphics{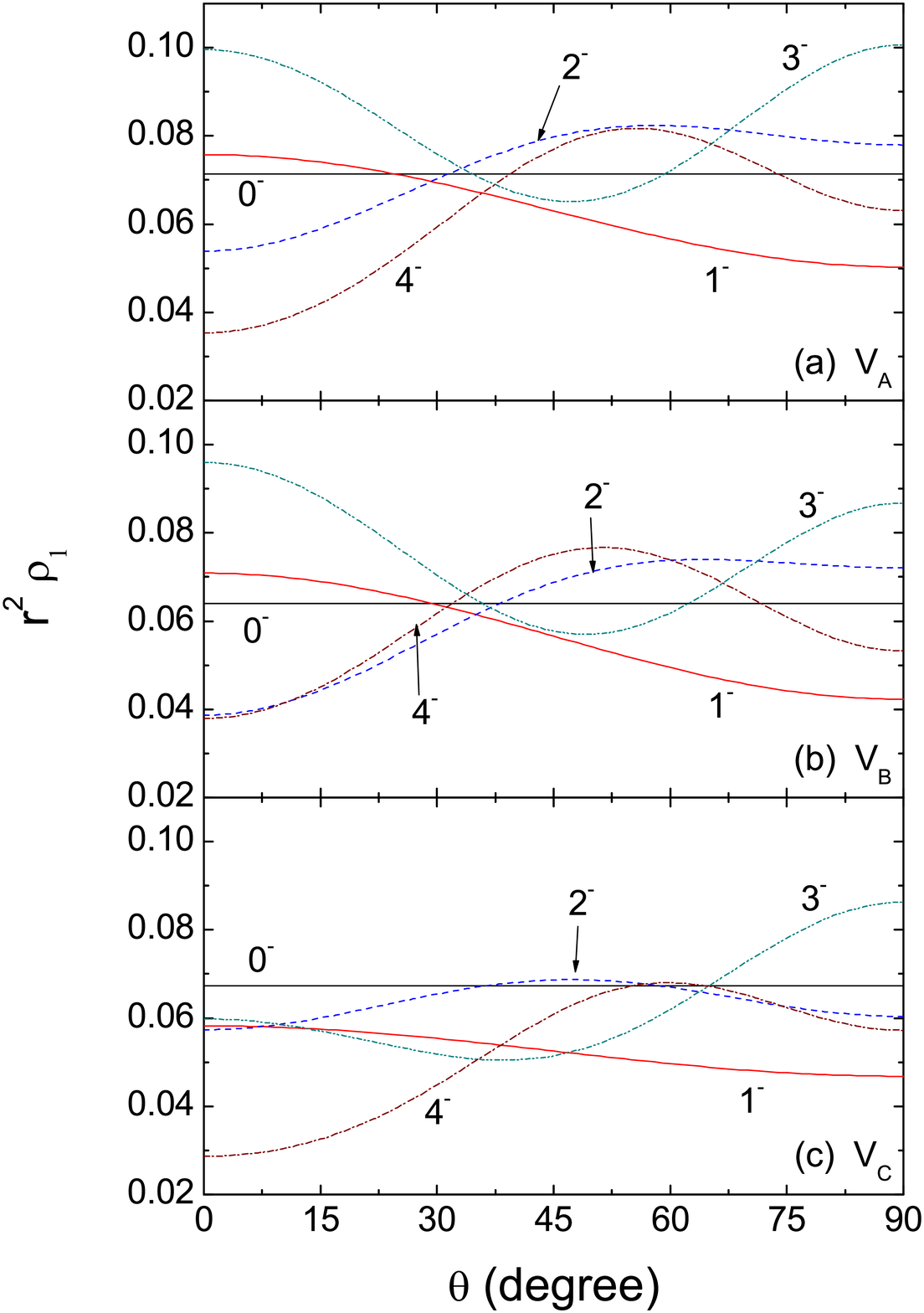} }
 \caption{Similar to Fig.~\ref{fig4} but for the odd parity
states $\Psi_{LL}^{(-)}$. }
 \label{fig5}
\end{figure}

The particle distributions in the $\Psi _{L,L}^{\Pi }$ states are
shown by the one-body densities $r^{2}\rho _{1}(r,\theta )$, where
$r$ and $\theta $ are the magnitude and the polar angle of a
position vector $\mathbf{r}$ originating from the c.m. (refer to
Appendix B). They are shown in Figs.~\ref{fig4} and ~\ref{fig5}. In
general the distribution depends on $M$. The choice $M=L$ implies
that the $Z $-axis is chosen to be lying along the direction of $L$.
In Fig.~\ref{fig4}, $0^{+}$ is isotropic as expected. $1^{+}$ is
nearly isotropic. $3^{+}$ and $4^{+}$ have a denser distribution
when the particle is closer to the equator ($\theta \rightarrow \pi
/2$). In general, when $L$ is larger, the system will more prefer to
increase its moment of inertia so as to reduce the rotation energy.
This explains why $3^{+}$ and $4^{+}$\ prefer to be a little
flattened. On the contrary, the distribution of $2^{+}$ \ prefers to
be closer to the $Z$-axis. This is not due to dynamics but due to
symmetry constraint. Since $2^{+}$\ is SQ-\textbf{ac} as shown in
Table~\ref{tab1}, it will pursue the SQ (This will be further
confirmed later). When the $\Sigma '$ as shown in Fig.~\ref{fig1} is
adopted, $\mathbf{k}'$ is chosen to be vertical to the SQ . However,
the SQ can only appear in $Q=0$\ component due to the constraint
Eq.~(\ref{e03_Psi}). In this component $L$\ is also nearly vertical
to $\mathbf{k}'$, and therefore $L$\ and the SQ are nearly coplanar.
Thus, in the choice $M=L$, SQ\ and the Z-axis are nearly coplanar.
This explains why the equator is less preferred by $2^{+}$.
Furthermore, 4b is strikingly similar to 4a, 4c is also similar to
4a but in a less extent. Thus, the angular distribution of particle
is not seriously affected by the details of interaction. In the
following, we shall see once and once that the similarity between
the cases with $V_{A}$ and $V_{B}$ is high, but weaker with $V_{C}$.
It has been mentioned that, for $V_{C}$, the pursuit to a better
geometry is less anxious. This cause the difference in the extent of
similarity.

High similarity between the cases $V_{A}$ and $V_{B}$ emerges
also in Fig.~\ref{fig5} for the odd-parity states.

\section{Shape-densities}

The similarity can be more clearly revealed by observing the
wave functions directly. For this purpose, we introduce the
hyper-radius $R\equiv \sqrt{\mu _{a}r_{a}^{2}+\mu
_{b}r_{b}^{2}+\mu _{c}r_{c}^{2}}$, where $\mu _{a}=\mu
_{b}=1/2$ and $\mu _{c}=1$ are the reduced masses. Note that
$R$ would remain invariant under the transformation between two
sets of Jacobi coordinates. Therefore $R$ is suitable for
measuring the size. From the normality and from
Eq.~(\ref{e02_Psi}), integrating over the Euler angles, we have
\begin{eqnarray}
 1
  =  \langle
     \Psi _{LM}^{\Pi }(X) |
     \Psi _{LM}^{\Pi }(X)
     \rangle
  = \frac{8\pi ^{2}}{2L+1}
    \sum_{Q}
    \int
    \mathrm{d}X' |
    \Psi _{LQ}^{\Pi }(X') |^{2}.
 \label{e08_1}
\end{eqnarray}
There are six arguments included in $X'$, one is $R$
responsible for the size and the others are denoted by $S$
responsible for the shape, and $\mathrm{d}X' =R^{8}\mathrm{d}R
\mathrm{d}S$. Thus the normality can be rewritten as $1=\int
\mathrm{d}R \mathrm{d}S\ \rho (X')$, where
\begin{eqnarray}
 \rho (X')
  \equiv
    \frac{8\pi ^{2}}{2L+1}
    R^{8}
    \sum_{Q} |
    \Psi_{LQ}^{\Pi }(X') |^{2},
 \label{e09_rho}
\end{eqnarray}
is the probability density that the system has a given size and
a given shape (while the orientation of the shape has already
been integrated). Therefore $\rho (X')$ is called the
shape-density.

\begin{figure}[tbp]
 \centering \resizebox{0.9\columnwidth}{!}{
 \includegraphics{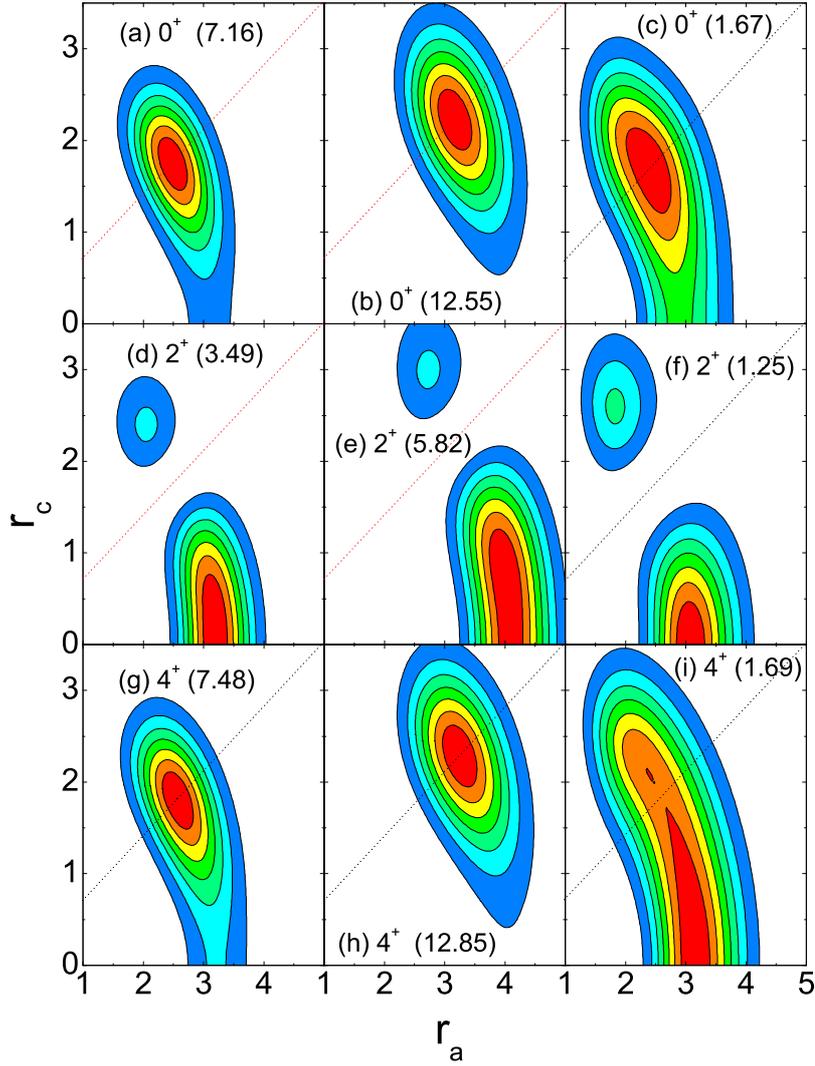} }
 \caption{The shape-densities $\rho(X')$ of even-parity states
observed in $Z_{2-2}$ (refer to Fig.~\ref{fig1}a) as a function of
$r_a$ and $r_c$. The three rows are for $0^+, 2^+,$ and $4^+$. The
three columns from left to right are for $V_A$, $V_B$, and $V_C$,
respectively (the same for Figs.~\ref{fig6} to \ref{fig9}). The
distribution of the density is relative to a body-frame, while the
orientation of the body-frame has been integrated. The dotted lines
have $r_c=r_a/\sqrt{2}$, they denote the ET shapes. In each panel,
the value of the maximum is given inside a pair of parentheses,
while the values of the contours form an arithmetic series and
decrease to zero. In Fig.6 - 9 the unit of length is $\sqrt{\hbar
/m\omega }$}
 \label{fig6}
\end{figure}

\begin{figure}[tbp]
 \centering \resizebox{0.9\columnwidth}{!}{
 \includegraphics{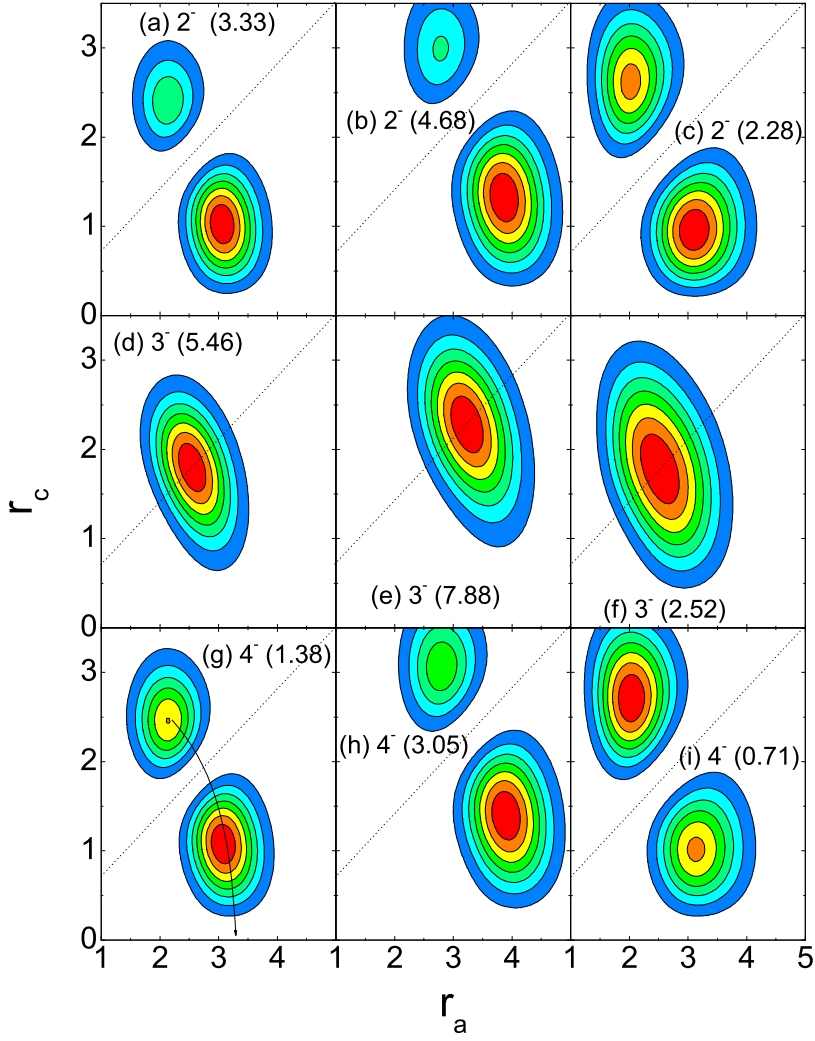} }
 \caption{The shape-densities $\rho(X')$ of odd-parity states
$2^-$, $3^-$, and $4^-$, respectively, observed in $Z_{2-2}$ as a
function of $r_a$ and $r_c$. Refer to Fig.~\ref{fig6}.}
 \label{fig7}
\end{figure}

We shall observe $\rho (X')$ in some selected zones. Since the
ET and the SQ are both contained in $Z_{2-2}$, this zone is
important. $\rho (X')$ of the even-parity and odd-parity states
observed in $Z_{2-2}$ are given in Figs.~\ref{fig6} and
\ref{fig7}, where $r_{a}$ and $r_{c}$ are variables. Since $
1^{+}$, $3^{+}$, $0^{-}$, and $1^{-}$ have their wave functions
absolutely being zero in this zone, they do not appear in these
figures (refer to Table~\ref{tab1}, where $Z_{2-2}$ is a PZ\
for them). Thus the states appear in the figures are $0^{+}$,
$2^{+}$, $4^{+}$, $2^{-}$, $3^{-}$, and $4^{-}$. Note that the
$ 0^{+}$ state is both ET-\textbf{ac} and SQ-\textbf{ac}. Due
to the superiority of the ET over the SQ, all the peaks in 6a,
6b, and 6c (for $V_{A}$, $V_{B}$, and $V_{C}$) are peaked at an
ET (where $r_{c}=r_{a}/\sqrt{2 }$) but not a SQ (where
$r_{c}=0$). It has been mentioned that, for $V_{C}$, the
pursuit to a better geometry is less anxious. Accordingly, the
distribution in 6c extends considerably toward a SQ . $4^{+}$
is also both ET-\textbf{ac} and SQ-\textbf{ac}. The SQ of the
$4^{+}$ can be nearly vertical to its $L$\ because $|Q|=4$\
component is allowed (refer to Eq.~(\ref{e03_Psi})). Thereby
the moment of inertia can be increased and the rotation energy
can be reduced. Accordingly, for $4^{+}$, the SQ becomes more
competitive. In fact, for $V_{C}$, the SQ has replaced the ET
as the most probable shape as shown in 6i.

$3^{-}$ is ET-\textbf{ac} but not SQ-\textbf{ac}. Accordingly,
the distribution is more concentrated around the ET and away
from the SQ as shown in Figs.~\ref{fig7}d, \ref{fig7}e, and
\ref{fig7}f.

$2^{+}$ is not ET-\textbf{ac} but SQ-\textbf{ac}. Accordingly,
the distribution is more concentrated around the SQ as shown in
Figs.~\ref{fig6}d, \ref{fig6}e, and \ref{fig6}f.

Among the four states of the second group, $2^{-}$ and $4^{-}$
are allowed to appear in this zone, and are shown in the first
and third rows of Fig.~\ref{fig7}. They are neither
ET-\textbf{ac} nor SQ-\textbf{ac}. There are two peaks in each
panel, the peak with a smaller $r_{c}$\ is the higher peak
(except in Fig.~\ref{fig7}i). It implies that the most probable
shape is a compressed ET along one of its two-fold axis.

 In general, one can define a specific degree of freedom
(accordingly, a specific 1-dimensional curve in the coordinate
space) to describe a specific mode of oscillation. When a wave
function is remarkably distributed along the curve then the specific
mode is involved. If a node appears at the curve, the oscillation
has been excited. If more nodes appear along the curve, the
excitation is more vigorous (the simplest example is referred to the
one-dimensional harmonic oscillator states). In each panel, let a
curve starting from the upper peak (a prolonged ET), extending to
the lower peak, and then extending to the abscissa (refer to the
solid curve added in Fig.~\ref{fig7}g). Accordingly, $r_{c}$
decreases along $\mathbf{k}$ (refer to Fig.~\ref{fig1}a) and the
prolonged ET suffers a compression and is transformed to a SQ. Once
$r_{c}$ becomes zero, this process will continue but in reverse
direction (namely, $r_{c}$ increases further along $-\mathbf{k)}$.
Then, the SQ is transformed back to a prolonged ET, and repeatedly.
The mode associated with the transformation is named the ET-SQ-ET
mode (or the H-type oscillation named in \cite{r_X4}). Since the
wave functions of $2^{-}$ and $4^{-}$ appear as nodes at the ET and
SQ, the ET-SQ-ET mode contains three nodes and therefore is an
excited mode. This explains why $2^{-}$ and $4^{-}$ are considerably
higher than $2^{+}$ and $4^{+}$.

From Table 1 we know that $Z_{1-3}$ is a PZ for $2^{-}$ but not for
$4^{-}$.  Thus the increase of rotation energy in $4^{-}$ can be
partially compensated due to the $Z_{1-3}$-\textbf{ac}. Therefore,
the energy difference between $2^{-} $ and $4^{-}$ is small. In
fact, the wave function of $4^{-}$ is also remarkably distributed in
$Z_{1-3}$. Say, for $V_{C}$, the maximum of $\rho (X')$ of $4^{-}$\
in $Z_{1-3}$ is equal to 1.52 located at a flattened regular
pyramid, whereas the maximum in $Z_{2-2}$ is equal to 0.71 located
at a prolonged ET (shown in Fig.~\ref{fig7}i). Thus, for $V_{C}$,
the most probable shape of $4 ^{-}$ is the flattened pyramid instead
of the prolonged ET.

\begin{figure}[tbp]
 \centering \resizebox{0.9\columnwidth}{!}{
 \includegraphics{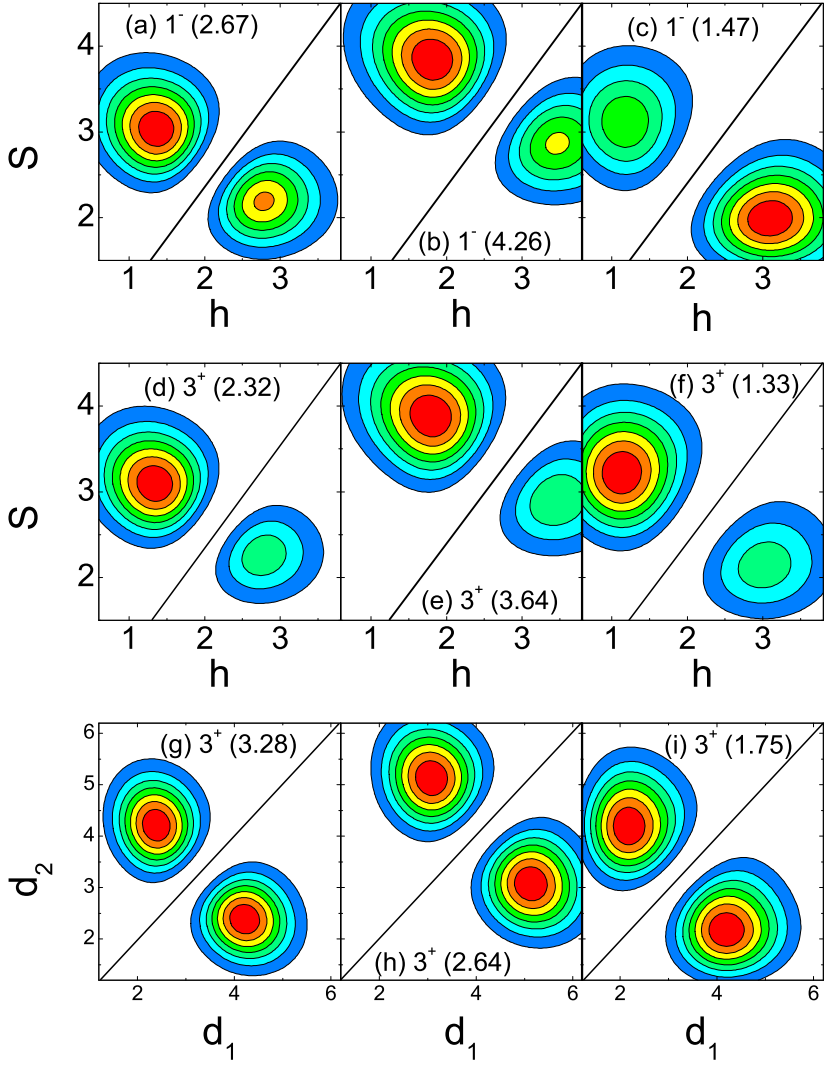} }
 \caption{ $\rho(X')$ of $1^-$ (a-c) and $3^+$ (d-f) plotted in
$Z_{1-3}$ for the regular pyramids. The height $h$ of the pyramid
and the side-length $s$ of the base triangle are variables. The
straight lines having $h=\sqrt{2/3}s$ are associated with an ET, and
they are an inherent nodal line for these states. $\rho(X')$ of
$3^+$ plotted in $D_{A\cap B}$ (g-i), where $d_1$ and $d_2$ are the
two diagonals of a rhombus. The straight line $d_1=d_2$ is
associated with the SQ, and it is an inherent nodal line for $3^+$.
}
 \label{fig8}
\end{figure}

The other two member $1^{-}$ and $3^{+}$ of the second group do not
appear in $Z_{2-2}$. From the spectra for $V_{A}$ and $V_{B}$, we
know that $1^{-}$ is higher than $2^{-}$, and $3^{+}$ is higher than
$4^{-}$. This fact implies that the prohibition of $Z_{2-2}$ will
lead to an increase in energy. From Table~\ref{tab1} we know that
both $1^{-}$ and $3^{+}$ will be distributed in $Z_{1-3}$. This is
shown in Fig.~\ref{fig8}a-\ref{fig8}c and \ref{fig8}d-\ref{fig8}f,
respectively. In all the panels, there are two peaks. The upper peak
has larger $s$ (i.e., the base triangle is larger), and is
associated with a flattened regular pyramid, while the lower peak a
prolonged pyramid. Let a curve be defined connecting the two peaks.
This curve is associated with the transformation between the
flattened and prolonged pyramid (or the K-type of oscillation
\cite{r_X4}). Since a node appears at the curve (more exactly, at an
ET), this oscillation has been excited. Also from Table~\ref{tab1}
we know that coplanar structures might exist in these two states.
This is shown in Figs.\ref{fig8}g-\ref{fig8}i, where the two peaks
are associated with a flattened and a prolonged diamond, and a node
appears between them. Thus an excited oscillation of a diamond
around a SQ is contained.

\begin{figure}[tbp]
 \centering \resizebox{0.9\columnwidth}{!}{
 \includegraphics{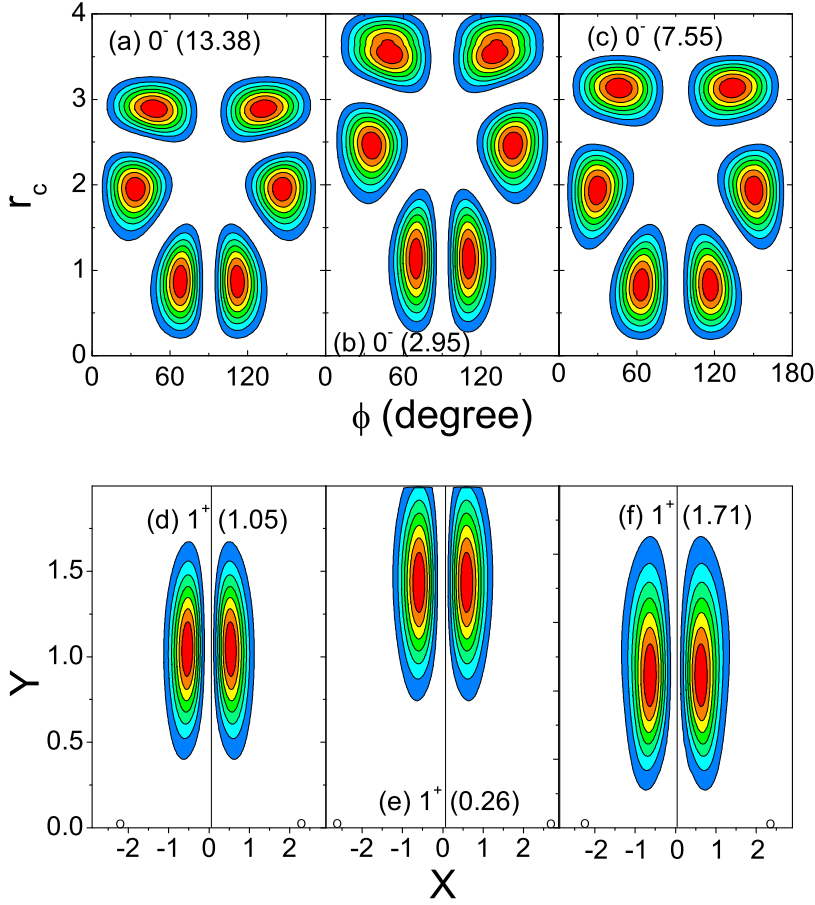} }
 \caption{(a-c) $\rho(X')$\ of $0^-$ plotted in a subspace of
$Z_{\mathrm{twi}}$ (refer to Fig.~\ref{fig1}b), where $r_c$ and
$\phi $ are variables, and $r_a = r_b$ depends on $r_c$ via a
constraint that the hyper-radius $R\equiv \sqrt{\mu
_{a}r_{a}^{2}+\mu _{b}r_{b}^{2}+ \mu _{c}r_{c}^{2}}$ is fixed at
$2r_{\mathrm{rms}}$ (p.s., for an ET, this constraint implies that
the distance of each particle from the c.m. is $r_{\mathrm{rms}}$ ).
(b) $\rho(X')$ of $1^+$ plotted in $D_{A}$ for the parallelograms.
This zone is a plane. The $X$- and $Y$-axes of a body frame with its
origin located at the c.m. is plotted. Two particles are given lying
at the X-axis marked by two small circles. The other two have
coordinates $(x,y)$ and $(-x,-y)$ . The peaks represent the most
probable locations of the third particle (while the fourth particle
at the lower plane can not be seen). Accordingly, the most probable
shape is a parallelogram. The vertical line at $x=0$ represents the
diamonds, and is an inherent nodal line for $1^{+}$. }
 \label{fig9}
\end{figure}

For the $0^{-}$ state of the third group any coplanar structures are
prohibited. This is due to the equivalence $R_{\pi
}^{\mathbf{k}'}\doteq I$, where $\mathbf{k}'$\ is chosen to be
vertical to the plane ($R_{\pi }^{\mathbf{k}'}$ causes no effect on
$L=0$\ states, while $I$\ causes a change of sign on odd parity
states, this leads to a contradiction) Thus, from Table~\ref{tab1},
the wave function of $0^{-}$ would be essentially distributed in
$Z_{\mathrm{twi}}$. However, many subspaces in $Z_{\mathrm{twi}}$
are PZ of $0^{-}$. This is shown in Figs.~\ref{fig9}a-\ref{fig9}c.
Where $\phi =0$ and $\pi $, and $r_{c}=0$ are\ three nodal lines
because they are associated with coplanar shapes (refer to
Fig.~\ref{fig1}b). $\phi =\pi /2$ is also a nodal line because the
associated shape belongs also to $Z_{2-2}$\ which is prohibited.
Furthermore, let the coordinates of the particle 2 relative to
$\Sigma '$\ be denoted as $(x,y,z)$. Accordingly, those of the
particles 1, 3, and 4 are $(-x,-y,z)$, $(-x,y,-z)$, and $(x,-y,-z)$,
respectively. One can see that, if $y=z$ the equivalence $R_{\pi
}^{\mathbf{j}'}IR_{\pi /2}^{\mathbf{i}'}\doteq P$ holds (where $P$
is an interchange of a pair of particles), if $z=x$ the equivalence
$R_{\pi}^{\mathbf{k}'}IR_{\pi /2}^{\mathbf{j}'}\doteq P$ holds, and
if x=y the equivalence $R_{\pi
}^{\mathbf{i}'}IR_{\pi/2}^{\mathbf{k}'}\doteq P$ holds. For $0^{-}$,
the left side of each of the equivalence will cause a change of sign
while the right side does not. Therefore $0^{-}$ must be zero in
each of the subspace with $y=z$, $z=x$, or $x=y$. The equality $y=z$
can be rewritten as $r_{c}/r_{a}=\sin (\phi /2)$. Note that $r_{c}$
and $r_{a}$ are constrained as
$\sqrt{r_{a}^{2}+r_{c}^{2}}=2r_{\mathrm{rms}}$\ in
Figs.~\ref{fig9}a-\ref{fig9}c. Thus $r_{c}=2r_{\mathrm{rms}}\sin
(\phi /2)/\sqrt{1+\sin ^{2}(\phi /2)}$ is a nodal line. Similarly,
the case z=x leads to the nodal line $r_{c}=2r_{\mathrm{rms}}\cos
(\phi /2)/\sqrt{1+\cos ^{2}(\phi /2)}$, while the case x=y leads to
the nodal line $\phi =\pi /2$ mentioned above. Totally, six nodal
lines are contained in Figs.~\ref{fig9}a-\ref{fig9}c. These lines
expel the wave function from favorable zones and cause strong
inherent oscillation. This explains why the energy of $0^{-}$ is so
high.

For $1^{+}$, due to the prohibition of $Z_{\mathrm{twi}}$ and
$Z_{1-3}$, the structure of tetrahedron is unfavorable.
Furthermore, due to the prohibition of $D_{B}$, the coplanar
structures are also unfavorable. It was found that this state
is a mixture of a parallelogram and a deformed tetrahedron. The
shape-density surrounding the parallelogram is shown in
Fig.s~\ref{fig9}d-\ref{fig9}f, where the vertical line is
associated with the diamonds and is a nodal line for $1^{+}$.
The two peaks (each peaked at a parallelogram) separated by the
nodal line implies an excited oscillation of the parallelogram.

From the shape-densities plotted in Figs.~\ref{fig6} to
\ref{fig9}, the decisive effect of the PZ is clearly shown.
Since the PZ is universal, the wave functions are expelled in a
universal way, and the room left for dynamic adjustment is
limited. This leads to similarity which is clearly shown by
comparing the three patterns of any row in the figures for
$V_{A}$, $V_{B}$, and $V_{C}$, respectively.

\section{Fermion systems}

The existence of PZ is not only for 4-boson systems but for all
kinds of few-body systems containing  identical particles. For an
example of a 4-fermion system, it is assumed that the spin-dependent
interaction tends to align the spins (alternatively, a magnetic
field is applied so that the spins are aligned). For this case, the
spatial wave functions will be fully anti-symmetric. Accordingly,
the equivalent operation Eq.~(\ref{e03_Psi}) that hold in $Z_{2-2}$
leads to the fact that $\Psi _{LQ}(X')$ is nonzero only if $Q=0,\pm
4,\cdots$ (when $\Pi =-1$), or $Q=\pm 2,\pm 6,\cdots$ (when $\Pi
=1$). In particular, the $Q=0$ component exists only if\ $\Pi =-1$
and $L$ being even. Accordingly, the ET-\textbf{ac} states are now
$0^{-}$, $3^{+}$, and $4^{-}$. The SQ-\textbf{ac} states are now
$2^{+}$, $3^{+}$, and $4^{+}$, while the states rejected from
$Z_{2-2}$ are now $0^{+}$, $1^{+}$, $1^{-}$, and $3^{-}$. Among them
$1^{+}$ and $3^{-}$ can both access $Z_{1-3}$, while $1^{+}$ can
access also coplanar structures. Therefore, for the ten states under
consideration, we can predict that $0^{-}$, $2^{+}$, $3^{+}$, and
$4^{-}$ would be the lowest, while $1^{-}$ and $0^{+}$ would be the
highest. We repeat our previous calculation for the case $V_{A}$\
but using fully anti-symmetric basis functions. The resultant
spectra is shown in Table~\ref{tab2}.

\begin{table}[htbp]
 \caption{The energies of the ten states (in ascending order) of
the fermion system calculated with $V_{\mathrm{A}}$ (the unit
is $\hbar \omega $).}
 \label{tab2}
 \begin{center}
  \begin{tabular}{c|cccccccccc}
   \hline\hline
   State  & $0^-$ & $3^+$ & $2^+$ & $4^-$ & $1^+$ & $2^-$ & $3^-$ & $4^+$ & $1^-$ & $0^+$ \\
   \hline
   Energy & 0     & 0.97  & 1.37  & 1.77  & 1.95  & 2.15  & 2.89  & 2.93  & 3.22  & 4.03 \\
   \hline\hline
  \end{tabular}
 \end{center}
\end{table}
The result in Table~\ref{tab2} confirms the prediction.

\section{Final remarks}

From the point of classical dynamics, it is incredible that
there are zones in the coordinate space favorable in energy but
absolutely prohibited. However, this paper demonstrates
analytically and numerically that this is a fact in microscopic
few-body systems. Since the prohibition arises from the
symmetry constraint which depends only on the inherent symmetry
but is irrelevant to dynamics, the existence of the PZ is
universal disregarding the kinds of systems and the details of
interaction. The quantum states of different systems will have
exactly the same PZ located at the same subspace when these
states are governed by the same inherent symmetry (namely, they
belong to the same set of representations of the symmetry
groups). Since some of the PZ have a higher geometric symmetry,
they are the zones more favorable in energy. Therefore, the
prohibition causes serious effect and destroys thoroughly the
picture painted simply by dynamics. Consequently, the
structures of the low-lying states are decisively determined by
their inherent symmetry, and the room left for the adjustment
from dynamics is narrow, in particular for the $i=1$ states

Although only 4-body systems with [1,1,1,1] and [$1^{4}$]
symmetries are concerned in this paper, obviously the existence
of the PZ is a universal phenomenon for all kinds of few-body
systems with identical particles.

\

\

Acknowledgment: The support from the NSFC under the grant
No.10874249 is appreciated.

\section*{Appendix}

\appendix

\section{Diagonalization of the Hamiltonian}

When the c.m. coordinates have been removed, the Hamiltonian
for internal motion $H_{\mathrm{int}}$ can be expressed by a
set of Jacobi coordinates
$\mathbf{r}_a=\mathbf{r}_2-\mathbf{r}_1$,
$\mathbf{r}_b=\mathbf{r}_4-\mathbf{r}_3$, and
$\mathbf{r}_c=(\mathbf{r}_3+\mathbf{r}_4-\mathbf{r}_1-\mathbf{r}_2)/2$.
Let
\begin{eqnarray}
 h(\mu,\mathbf{r})
  \equiv
    -\frac{1}{2\mu}
     \nabla_{\mathbf{r}}^2
    +\frac{1}{2}
     \mu
     r^2,
 \label{e_A1}
\end{eqnarray}
which is the Hamiltonian of harmonic oscillation, Then
\begin{eqnarray}
 H_{\mathrm{int}}
  =  h(1/2,\mathbf{r}_a)
    +h(1/2,\mathbf{r}_b)
    +h(1,\mathbf{r}_c)
    +\sum_{i<j}
    V(|\mathbf{r}_j-\mathbf{r}_i|).
 \label{e_A2}
\end{eqnarray}

Let us introduce a variational parameter $\gamma$. The
eigenstates of $h(\gamma,\mathbf{r})$ are denoted as
$\phi_{nl}^{\gamma}(\mathbf{r})$, where $n$ and $l$ are the
radial and angular quantum numbers, respectively. From
$\phi_{nl}^{\gamma}(\mathbf{r})$ a set of basis functions for
the 4-body system is defined as
\begin{eqnarray}
 \Phi_{k,\Pi LM}^{\gamma}(1234)
  \equiv
     [ ( \phi_{n_a l_a}^{\gamma/2}(\mathbf{r}_a)
         \phi_{n_b l_b}^{\gamma/2}(\mathbf{r}_b) )_{l_{ab}}
       \phi_{n_c l_c}^{\gamma}(\mathbf{r}_c) ]_{LM},
 \label{e_A3}
\end{eqnarray}
where $l_a$ and $l_b$ are coupled to $l_{ab}$, then $l_{ab}$
and $l_c$ are coupled to the total orbital angular momentum $L$
and its $Z$-component $M$, $\Pi$ the parity, $k$ denotes the
set of quantum numbers. These functions should be further
symmetrized. For bosons, $l_a$ and $l_b$ should be even and we
define
\begin{eqnarray}
 \tilde{\Phi}_{k,\Pi LM}^{\gamma}
  \equiv
     \sum_p
     \Phi_{k,\Pi LM}^{\gamma}(p_1 p_2 p_3 p_4),
 \label{e_A4}
\end{eqnarray}
where the right side is a summation over the permutations.
Making use of the Talmi-Moshinsky coefficients,
\cite{r_BTA,r_BM,r_TW} each term at the right can be expanded
in terms of $\Phi_k^{\gamma}(1234)$. For an example,
\begin{eqnarray}
 \Phi_{k,\Pi LM}^{\gamma}(1324)
  =  \sum_{k'} A_{kk'}^{\Pi L}
     \Phi_{k',\Pi LM}^{\gamma}(1234),
 \label{e_A5}
\end{eqnarray}
where $A_{kk'}^{\Pi L}$ is the Talmi-Moshinsky coefficients
which can be obtained by using the method given in the
ref.\cite{r_TW}. Note that the set $\{\tilde{\Phi}_{k,\Pi
LM}^{\gamma}\}$ has not yet been orthonormalized. Thus a
standard procedure is needed to perform to transform
$\{\tilde{\Phi}_{k,\Pi LM}^{\gamma}\}$ to a orthonormalized set
$\{\tilde{\tilde{\Phi}}_{q,\Pi LM}^{\gamma}\}$, where $q$ is a
serial number to denote a basis function, and each basis
function can be expanded in terms of $\{\Phi_{k,\Pi
LM}^{\gamma}(1234)\}$. Finally, the set
$\{\tilde{\tilde{\Phi}}_{q,\Pi LM}^{\gamma}\}$ is used to
diagonalize $H_{\mathrm{int}}$. Due to the Talmi-Moshinsky
coefficients, only the set of coordinates $\mathbf{r}_a$,
$\mathbf{r}_b$, and $\mathbf{r}_c$ is involved. Thus the
calculation of the matrix elements is straight forward. When
$\Pi$ and $L$ are given, a series of states will be obtained
after the diagonalization. The lowest one of the series is
called the first state. $\gamma$ has to be adjusted so that the
energy of the first state is as low as possible.

Due to having a trap, the bound states will tend to zero quite
fast when $r$ tends to infinity. For this case, the above basis
functions are suitable and will lead to a better convergency.
Let $N_{abc}=2(n_{a}+n_{b}+n_{c})+l_{a}+l_{b}+l_{c}$, and let
the number of basis functions be constrained by $N_{abc}\leq
N_{max}$. Then, as an example of the bosonic case, the
resultant energy of the lowest $0^{+}$ state with the
interaction $V_{A}$ is 2.2965, 2.2961, and 2.2960 $\hbar \omega
$, respectively, when $N_{max}= 16$, 18, and 20. Since the
emphasis is placed at the qualitative aspect, the convergency
appears to be satisfactory.

\section{One-body density $r^{2}\rho _{1}$ , the root mean square
radius $r_{\mathrm{rms}}$}

After the diagonalization, each eigenstate $\Psi_{L,M}^{\Pi,i}$
is expanded in terms of $\tilde{\tilde{\Phi}}_{q,\Pi
LM}^{\gamma}$, where $i$ denotes the $i$-th state of the $(\Pi
L)$-series. For convenience, a transformation from the set
${\tilde{\tilde{\Phi}}_{q,\Pi LM}^{\gamma}\}}$ to the set
$\{\Phi_{k,\Pi LM}^{\gamma}(1234)\}$ is made, and the
eigenstate is re-expanded as $\Psi_{L,M}^{\Pi,i}=\sum_k
C_k^{\gamma i\Pi L}\Phi_{k,\Pi LM}^{\gamma}(1234)$. In order to
extract the information of a particle from
$\Psi_{L,M}^{\Pi,i}$, another set of Jacobi coordinates
$\mathbf{r}_a=\mathbf{r}_2-\mathbf{r}_1$,
$\mathbf{r}_d=(\mathbf{r}_1+\mathbf{r}_2)/2-\mathbf{r}_4$, and
$\mathbf{r}_e=\mathbf{r}_3-(\mathbf{r}_4+\mathbf{r}_1+\mathbf{r}_2)/3$
is defined. Accordingly, we have the basis function
\begin{eqnarray}
 \Upsilon_{k,\Pi LM}^{\gamma}(1234)
  \equiv
     [ ( \phi_{n_a l_a}^{\gamma/2}(\mathbf{r}_a)
         \phi_{n_b l_b}^{2\gamma/3}(\mathbf{r}_d) )_{l_{ab}}
       \phi_{n_c l_c}^{3\gamma/4}(\mathbf{r}_e)]_{LM}.
 \label{e_A6}
\end{eqnarray}

Since the set $\{\Phi_{k,\Pi LM}^{\gamma}(1234)\}$ and the set
$\{\Upsilon_{k,\Pi LM}^{\gamma}(1234)\}$ can be transformed
into each other via the Talmi-Moshinsky coefficients, the
eigenstate can once again expanded as
\begin{eqnarray}
 \Psi_{L,M}^{\Pi,i}
  =  \sum_k
     D_k^{\gamma i\Pi L}
     \Upsilon_{k,\Pi LM}^{\gamma}(1234),
 \label{e_A7}
\end{eqnarray}
where the coefficients $D_k^{\gamma i\Pi L}$ can be known from
the diagonalization and from the transformation between sets of
basis functions.

Now, from the normality, we have
\begin{eqnarray}
 1
 &=& \int
     \mathrm{d}\mathbf{r}_e
     \mathrm{d}\mathbf{r}_a
     \mathrm{d}\mathbf{r}_d
     (\Psi_{L,M}^{\Pi,i})^*
     \Psi_{L,M}^{\Pi,i}  \nonumber \\
 &=& \int
     \mathrm{d}r\
     \sin\theta
     \mathrm{d}\theta \nonumber \\
  &&
     [ 2\pi
      ( \frac{4}{3})^3
        r^2
        \int
        \mathrm{d}\mathbf{r}_a
        \mathrm{d}\mathbf{r}_d
        \sum_{k',k}
        D_{k'}^{\gamma i\Pi L}
        D_k^{\gamma i\Pi L}(\Upsilon_{k',\Pi LM}^{\gamma}(1234))^*
        \Upsilon_{k,\Pi LM}^{\gamma}(1234) ],
 \label{e_A8}
\end{eqnarray}
where $\theta$ is the polar angle of $\mathbf{r}_e$, and
$r=\frac{3}{4}r_e$ is the distance of particle 3 from the c.m.
Obviously, the quantity inside the bracket is just the one-body
density $r^2\rho_1(r,\theta)$ describing the particle
distribution (all the four particles are distributed in the
same way).

The root mean square radius is
\begin{eqnarray}
 r_{\mathrm{rms}}
  =  \frac{3}{4}
     ( \sum_{k',k}
       D_{k'}^{\gamma i\Pi L}
       D_{k}^{\gamma i\Pi L}
       \langle
       \Upsilon _{k',\Pi LM}^{\gamma}(1234) |
       r_{e}^{2} |
       \Upsilon _{k,\Pi LM}^{\gamma }(1234)
       \rangle )^{1/2}.
 \label{e_A9}
\end{eqnarray}

\end{document}